\documentclass[aip, apl, 10pt, twocolumn,unsortedaddress]{revtex4-1}
\usepackage{graphicx}
\usepackage{epstopdf}

\begin{document}

\title{Dispersionless optical activity in metamaterials}

\author{Kirsty Hannam}
\email{kirsty.hannam@anu.edu.au}
\author{David A. Powell}
\author{Ilya V. Shadrivov}
\author{Yuri S. Kivshar}

\affiliation{Nonlinear Physics Centre, Research School of Physics and Engineering, Australian National University, Canberra, ACT 0200, Australia}

\begin{abstract}
We introduce a chiral metamaterial with strong, non-resonant optical activity, and very low polarization ellipticity. We achieve this by combining a meta-atom and its complementary structure into a meta-molecule, resulting in the coupling of magnetic and electric dipole responses. In contrast to either a pair of crosses, or complementary crosses, this structure has low dispersion in the optical activity at the transmission resonance. We also study the excitation mechanism in this structure, and optimize the optical activity through changing the twist angle.
\end{abstract}

\maketitle

A chiral structure is distinct from its mirror image, which causes the degeneracy between the right- and left-handed circularly polarized waves in the structure to be broken. This is due to cross-coupling between the magnetic and electric polarizations of the media at resonance, resulting in optical activity and circular dichroism~\cite{Lietal2008}. The response in metmaterials is much stronger than that in natural materials.

Three-dimensional structures, such as the helix, or canonical spiral, can result in coupled magnetic and electric dipole type responses, due to the currents around the loops, and along the structure~\cite{Saenzetal2008}. However, three-dimensional structures are difficult to fabricate, especially when scaled down for use at terahertz and optical frequencies.

Alternatively, by combining two or more achiral, planar elements, such as crosses~\cite{Deckeretal2009} or split ring resonators (SRRs)~\cite{Deckeretal2010}, rotated about their common axis, a chiral meta-atom can be created. Such configurations have been shown to exhibit strong optical activity, and broadband polarization conversion~\cite{Deckeretal2009, Deckeretal2010,Weietal2011b, Zhaoetal2012}. It is known that the strong, near-field response between neighboring meta-atoms is important in determining the properties of the metamaterial~\cite{Powelletal2010, Powelletal2011}. This is also the case in chiral structures formed from achiral constituents, where the near-field interaction is essential for determining the overall response of the material~\cite{Hendryetal2012, Zhaoetal2012}. The coupling effects between multiple such chiral structures have also been studied~\cite{Andryieuskietal2010,Lietal2010}.

Depending on the resonant mode, the response of these structures is dominated by their electric or magnetic dipole moment. In either case, this results in strong reflection at the resonant frequency due to impedance mismatch of the sample with the surrounding medium. In addition, the resonant optical activity in such structures is accompanied by strong circular dichroism, causing ellipticity of the output polarization state, which is often undesirable~\cite{Deckeretal2009}. In conducting a thorough search of the relevant literature~\cite{Deckeretal2009,Weietal2011b,Deckeretal2010,Hendryetal2012, Zhaoetal2012,Lietal2011,Andryieuskietal2010,Lietal2010, Dingetal2012,Lietal2012}, it is found that for all relevant cases this dispersive optical activity occurs. It is possible to achieve reasonably flat optical activity off-resonance~\cite{Deckeretal2009}, however this is accompanied by a drop-off in the magnitude of the optical activity.

By combining a meta-atom with its complement in a chiral configuration, we can overcome the impedance mismatch at the resonant frequency, as according to Babinet's principle this results in the coupling of an electric dipole type response with a corresponding magnetic dipole type response~\cite{Falconeetal2004}. This results in an impedance matching effect and low reflection at resonance, and should also lead to higher transmission. This approach has been used to achieve dual-band ultraslow modes, by alternating layers of SRRs and their complementary structures~\cite{NavarroCiaetal2010}. It has also been used to create a broad bandpass filter in the terahertz regime, combining a cross and it's complement of different parameters in a non-chiral arrangement, for which the current distributions for the resonant modes were also studied~\cite{Chiangetal2011}. 

Broadband quarter-wave plate operation was achieved recently by tailoring the resonances of perpendicular nanorods~\cite{Zhao&Alu2013}. However, this system was operated off-resonance, and did not involve close coupling of the resonant elements, unlike our system.

Here we study a cross coupled to its complement, or a ``mixed pair'', and find that our structure provides strong optical activity which resonates away from the transmission resonance, resulting in very low ellipticity at the transmission resonance. We also determine, by looking at the case of a strip combined with a slot, that our structure is excited by means of the hole-modes in the complementary cross. We optimize the chirality through the twist angle, and study the effect of changing the spacing between elements.

\begin{figure}[tb]
	\centering
		\includegraphics[width=\columnwidth]{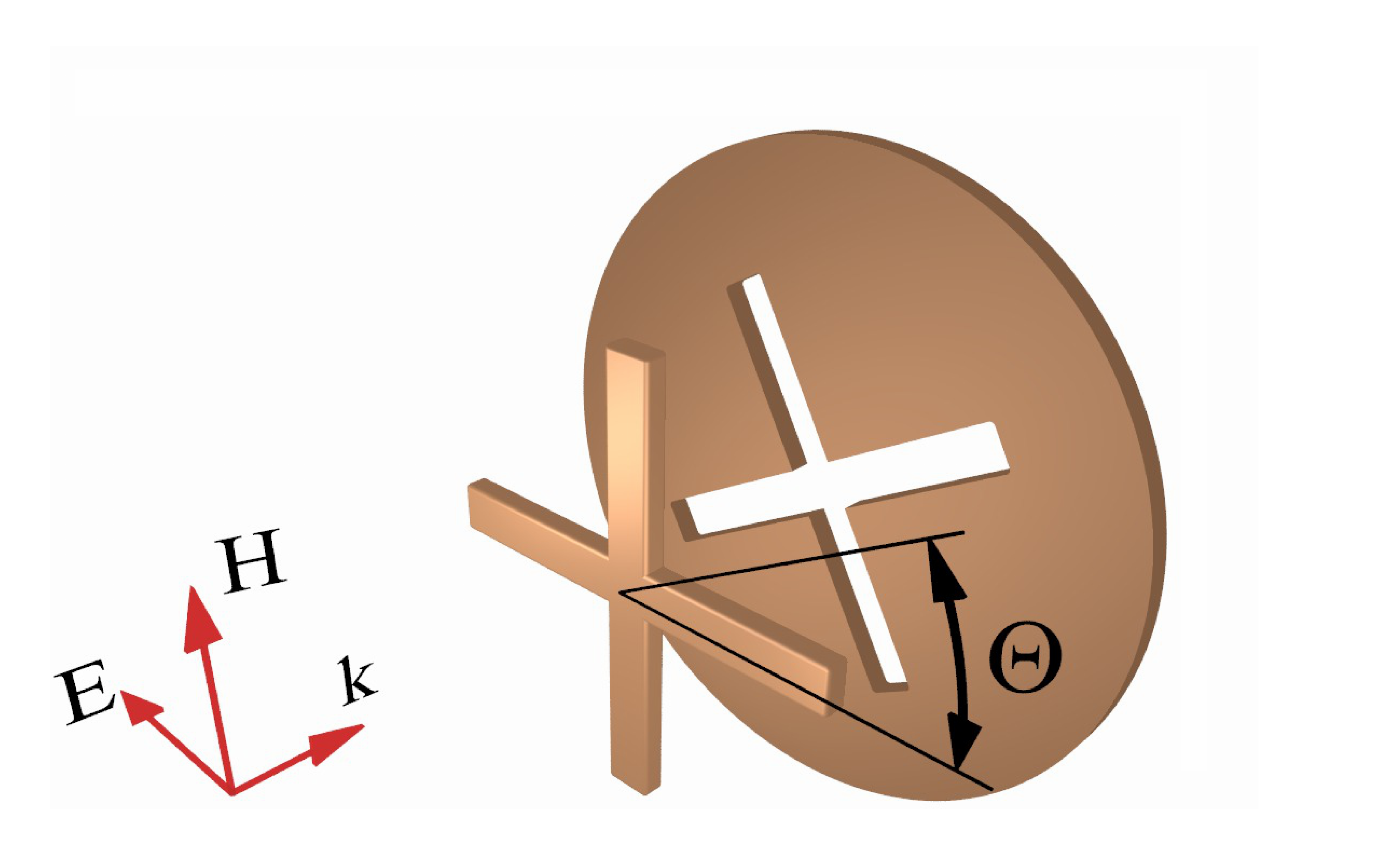}
	\caption{Schematic of our proposed mixed structure: a cross coupled to its complement, rotated through an angle $\theta$.}
	\label{fig:Schem}
\end{figure}

We choose the cross and its complement to have arms of length $25$mm, and width $1.5$mm. They are separated by a substrate $1.6$mm thick, and rotated through $22.5^{\circ}$. We model them as perfect electrical conductors (PEC), inside a circular waveguide, so as to be experimentally realizable. The substrate has a dielectric constant $4.3$ and loss tangent of $2.5$x$10^{-4}$.  A schematic of the two elements rotated through an angle $\theta$ is shown in Fig.~\ref{fig:Schem}. Simulations are performed using CST Microwave Studio, using a linearly polarized input wave, where the first two cut-off modes are excited. The first mode is assigned to that with the electric field oriented in the x-direction, and the second for the y-direction. 

\begin{figure}[tb]
	\centering
		\includegraphics[width=0.9\columnwidth]{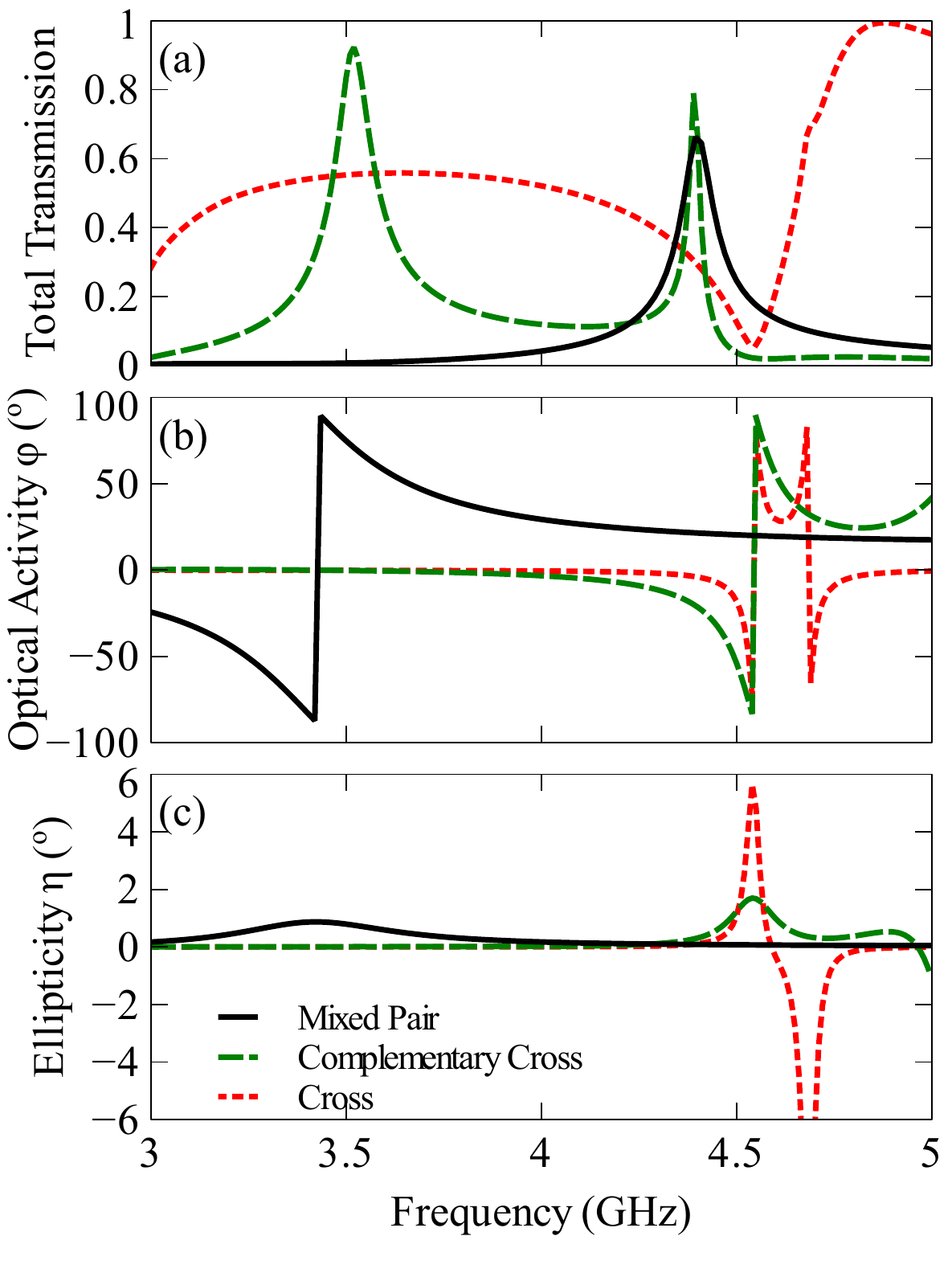}
	\caption{(a) Transmission, (b) optical activity and (c) ellipticity for two crosses, two complementary crosses and the mixed pair.}
	\label{fig:1pair}
\end{figure}

We simulate the co- and cross-polarized transmission coefficients for both linear polarizations ($S_{xx}$, $S_{yy}$, $S_{xy}$ and $S_{yx}$). As our structure has four-fold rotational symmetry, we only need $S_{xx}$ and $S_{xy}$. The total transmission amplitude ($S_{xx}^2 + S_{xy}^2)^{1/2}$ is plotted in Fig.~\ref{fig:1pair}(a). 

Transmission coefficients for circular polarization are found as
\begin{equation}
T^{\pm} =\frac{((S_{xx} + S_{yy}) \mp i(S_{xy} - S_{yx}))}{2},
\end{equation}
then from this we calculate the optical activity as
\begin{equation}
\phi = \frac{arg(T^{++}) - arg(T^{--}) + 2m\pi}{2},
\label{eq:oa}
\end{equation}
where $m$ is an integer such that $\phi$ is between $-\pi$ and $\pi$. This is plotted in Fig.~\ref{fig:1pair}(b) (solid black curve).  We show the corresponding ellipticity, in Fig.~\ref{fig:1pair}(c) (solid black curve), which is calculated as

\begin{equation}
\eta = \frac{1}{2}\tan^{-1}\frac{|T^{++}|^{2}-|T^{--}|^{2}}{|T^{++}|^{2}+|T^{--}|^2}.
\label{eq:E}
\end{equation}

For comparison's sake, we also calculate the optical activity and ellipticity for a pair of crosses, and a pair of complementary crosses, both also with an internal rotation of $\theta = 22.5^{\circ}$. The lengths of the elements in these structures were designed so that the transmission resonances line up with those of the mixed structure -  ($24$mm) $16$mm for the (complementary) crosses, [Figs~\ref{fig:1pair}(a, b)].

In Fig.~\ref{fig:1pair}(a) we see a resonant pass band for the mixed structure at $4.4$GHz. There is also a stop band at $3.4$GHz, which cannot be seen in this plot as the background transmission is already low. For the pair of crosses, we see a resonant stop band at $4.6$GHz. We see two resonant pass bands for the pair of complementary crosses, at $3.5$ and $4.4$GHz. By reducing the distance between the two crosses, thus increasing the coupling, we find that there are two resonances. The mixed structure has at least one other resonance below the cut-off frequency of the waveguide, however the pass band at $4.4$GHz is the most useful resonance, as it is a pass band accompanied by large optical activity.

In Fig.~\ref{fig:1pair}(b) we compare the optical activities for the three structures, as calculated using Eq.~(\ref{eq:oa}). We see that both the pair of crosses and pair of complementary crosses show highly dispersive optical activity at the resonant transmission band. For the mixed structure, we see a resonance in the optical activity at $3.4$ GHz. This resonance is caused by the stop band, where the through transmission dips below the cross-polarized transmission. The optical activity at the pass band frequency has very low dispersion, but is still large (about $20^{\circ}$). 

The corresponding ellipticities, calculated using Eq.(~\ref{eq:E}), are shown in Fig.~\ref{fig:1pair}(c). The ellipticity corresponds to the gradient of the optical activity shown in Fig.~\ref{fig:1pair}(b). This means that, unlike the other two structures, our mixed structure has very low ellipticity at resonance,  which is very desirable. It also shows much lower ellipticity overall than the other structures.

\begin{figure}[tb]
	\centering
		\includegraphics[width=\columnwidth]{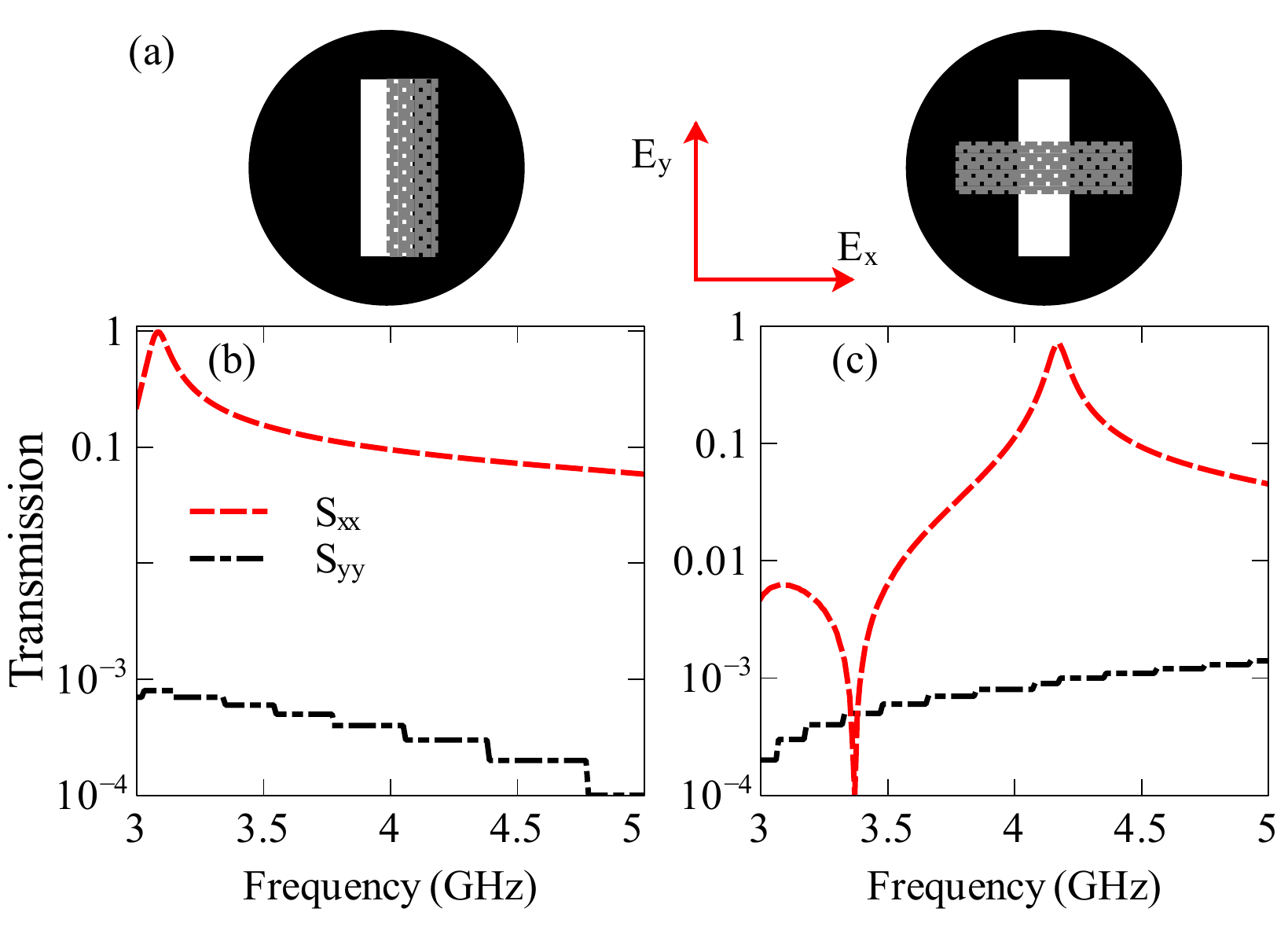}
	\caption{(a) Basic schematics of the strip, and slot, in parallel, and perpendicular arrangements. The slot is aligned with the y-axis, the strip is rotated. The gray, hatched rectangle is the strip and the black, solid outline is the slot. The simulated transmissions for the both modes are shown in (b) (parallel) and (c) (perpendicular).}
	\label{fig:Strips}
\end{figure}

In order to determine the nature of the transmission resonances of this structure, we study the excitation of a single strip and slot. We choose to do this as the slot-strip system is a simpler structure than our ``mixed pair", and is also anisotropic, but exhibits qualitatively similar behavior as our structure. It allows us to determine which coupling mechanisms contribute to the response in our system. We align the slot along the y-axis, then add the strip aligned either in a parallel or perpendicular configuration, as shown in Fig.~\ref{fig:Strips}(a). We use strips of the same length and width as the crosses above, along with the same substrate, resulting in a transmission peak in the same frequency range. Fig.~\ref{fig:Strips}(b) shows the through transmission for both incoming polarizations, for the parallel configuration of this set-up. The same results, but for the perpendicular arrangement, are shown in Fig.~\ref{fig:Strips}(c).  

We see that for both arrangements, in Figs~\ref{fig:Strips}(b,c), significant transmission is only seen when the incoming polarization is across the slot ($S_{xx}$). It should be noted that these graphs are plotted using a log scale, so it can be seen that the transmission for the wave polarized along the slot ($S_{yy}$) is almost negligible. By looking at these results, we can conclude that the predominant mechanism in exciting this structure is through the hole-mode in the slot. The strip is excited by the electric field parallel to it. As the strip is rotated from parallel to perpendicular to the slot, the strength of its excitation changes. This then couples to the excitation of the slot to determine the properties of the resonance.

We can then conclude that the predominant excitation in our structure is through the hole-mode in the complementary structure aligned perpendicular to the magnetic field, which then couples primarily to the perpendicular corresponding arm of the cross.

\begin{figure}[tb]
	\centering
		\includegraphics[width=\columnwidth]{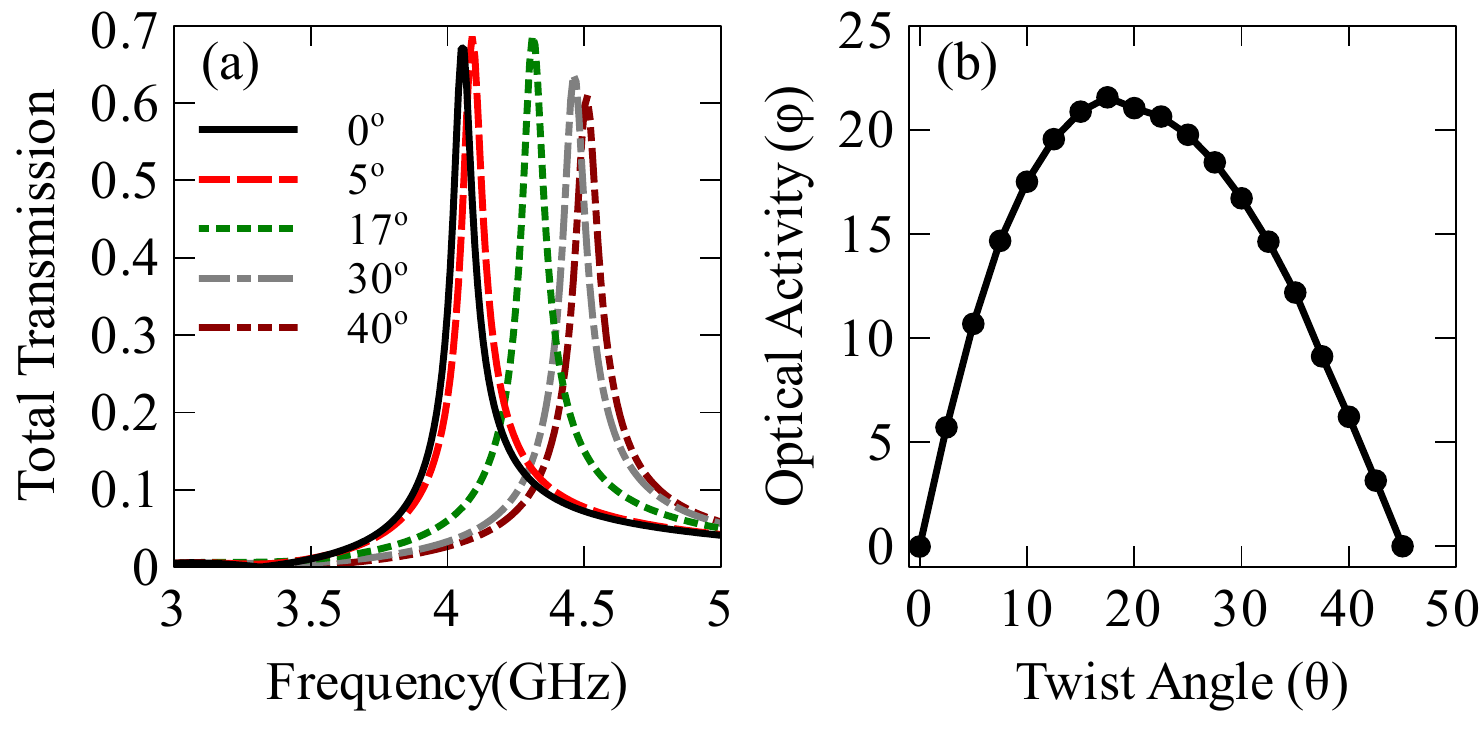}
	\caption{(a) Total transmission for various twist angles $\theta$; and (b) Maximum optical activity at resonance as a function of twist angle.}
	\label{fig:Angle}
\end{figure}

As the transmission resonances are similar in the chiral and achiral configurations, we now study the effect of introducing chirality through the twist angle $\theta$. We measure the transmission for $\theta$ ranging from $\theta = 0^{\circ}$ to $45^{\circ}$, in $2.5^{\circ}$ increments. In Fig.~\ref{fig:Angle}(a) we show the total transmission for a few of these angles. We see that as $\theta$ is increased, the resonance increases, and the transmission height changes slightly, but the effect of $\theta$ on the transmission is not huge. Fig.~\ref{fig:Angle}(b) shows the optical activity at the resonance frequency, as a function of twist angle $\theta$. We see that the near-field coupling within a twisted ``mixed pair" leads to changing optical activity due to change in the coupling between the two elements.  While this structure has maximum asymmetry at $\theta = 22.5^{\circ}$, we find maximum optical activity of $\phi = 22^{\circ}$ at $\theta =17.5^{\circ}$. This is due to retardation, as shown in Ref.~\onlinecite{MLiuetal2012a}. The resonant behavior at $3.5$GHz, associated with the stop-band resonance in the transmission, is present for most values of $\theta$, and is less tunable. The ellipticity at resonance was also calculated, and shows a similar trend (not shown here), peaking at $0.1^{\circ}$.

\begin{figure*}[t]
	\centering
		\includegraphics[width=0.9\textwidth]{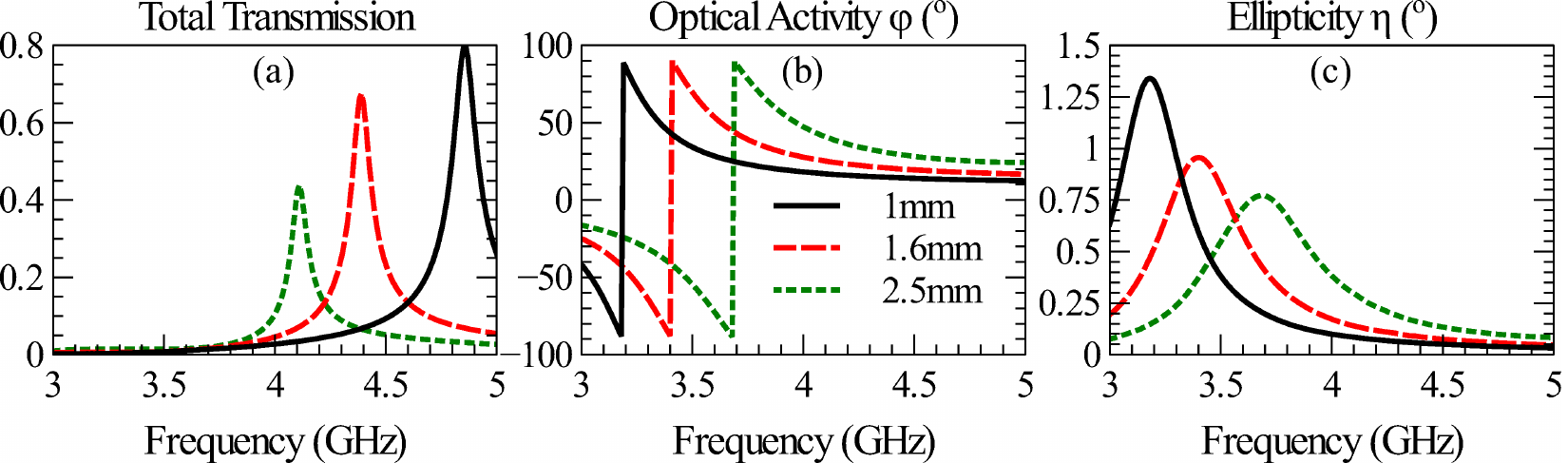}
	\caption{(a) Total transmission, (b) optical activity and (c) ellipticity for various thicknesses of the dielectric.}
	\label{fig:Distance}
\end{figure*}

By changing the distance between the cross and its complement, we change the retardation in the structure, as well as the interaction between the resonators. From Fig.~\ref{fig:Distance}(a) we see that by increasing the spacing between the elements, we significantly reduce the magnitude of the transmission. More importantly, the transmission resonance shifts closer to the optical activity resonance, therefore increasing the ellipticity of the structure across the transmission band. However, we also see a decrease in the magnitude of the optical activity as the spacing is decreased. Therefore we conclude that we have a trade-off between the magnitudes of the transmission and optical activity, in choosing the optimal spacing. 

In conclusion, we have proposed a ``mixed pair" structure, which is a combination of a meta-atom with its complement, and found large, flat optical activity at resonance, accompanied by very low ellipticity. We have also shown how this structure is excited, and how these effects can be optimized by changing the twist angle $\theta$, and the spacing between the elements. 

%

%\bibliographystyle{aipnum4-1}
%\bibliography{Tuning}

\end{document}